# AIDES EN LIGNE À L'UTILISATION DE LOGICIELS GRAND PUBLIC : PROBLEMES SPECIFIQUES DE CONCEPTION ET SOLUTIONS POTENTIELLES

Antonio Capobianco, Noëlle Carbonell

**Résumé**

On peut faire aujourd'hui le même constat qu'il y a vingt ans, à savoir que les aides en ligne à l'utilisation des logiciels grand public sont rarement consultées par les utilisateurs novices. Pourtant, si les recherches sur l'aide en ligne suscitent moins d'intérêt depuis quelques années, les systèmes d'aide à l'utilisation des logiciels grand public commercialisés ont évolué considérablement au cours des vingt dernières années.

Cet article prend d'abord position en faveur de la nécessité d'une assistance en ligne à l'utilisation d'un nouveau logiciel, dans le débat de fond qui divise la communauté scientifique en interaction homme-machine sur la question fondamentale suivante : les aides en ligne sont-elles un pis-aller, un simple remède aux défauts actuels de conception des logiciels et des interfaces utilisateur, ou bien sont-elles une assistance d'une nécessité absolue pour acquérir la maîtrise de l'utilisation d'un nouveau logiciel ? La discussion s'appuie à la fois sur les résultats d'études empiriques ou expérimentales et sur des arguments théoriques. La seconde partie analyse les obstacles et les difficultés spécifiques auxquels se heurte la conception d'aides en ligne à l'intention du grand public, afin de comprendre pourquoi les aides en ligne sont ignorées des utilisateurs en dépit de leur nécessité. Dans la dernière partie, les contributions potentielles de diverses approches à la résolution de ces problèmes spécifiques de conception sont analysées et discutées. Ces approches mettent en œuvre des paradigmes et techniques d'interaction récents, dont la personnalisation statique et dynamique de l'interaction, l'aide en ligne contextuelle et de nouvelles formes de multimodalité qui intègrent la parole, en entrée et en sortie, aux modalités classiques d'interaction.

**Abstract**

Despite long-standing research efforts, online help systems are unable yet to satisfy the needs of novice users in the general public, in-as-much as these users tend to ignore them.
This paper first demonstrates the necessity of online help for getting familiar with the use of new software intuitive though it might be. The main issue addressed is whether online help systems are meant to make up for design weaknesses or to meet novices' intrinsic needs. The discussion involves both empirical results and theoretical arguments.
In the second part, major specific difficulties that online help designers are confronted with are analysed with a view to understanding why available online help systems are underused by novices although they are essential.

Dès l'apparition des ordinateurs personnels (PC) au début des années 80, l'aide en ligne à l'intention des utilisateurs non informaticiens, notamment du grand public, a suscité de nombreuses recherches en ergonomie des logiciels, et fait l'objet de débats et de controverses scientifiques qui ont favorisé une diversification des approches et stimulé les efforts des chercheurs. Cependant, depuis le milieu des années 90, les recherches se sont raréfiées en dépit des besoins et d'une demande qui ne cessent de croître. En effet, le nombre des utilisateurs est en constante augmentation. Par exemple, Netcraft[1] a observé, au cours du mois de septembre 2006, une augmentation du nombre de sites Web de l'ordre d'un million. Cet organisme évalue le nombre de sites Web actifs en octobre 2006, à près de 50 millions. En outre, les logiciels grand public se diversifient et les versions successives d'un même logiciel se succèdent à un rythme qui s'accélère, et avec des évolutions de l'application et de l'interface utilisateur importantes. D'une version à la suivante, l'enrichissement des fonctionnalités et l'augmentation de la flexibilité de l'interface se traduisent souvent par un accroissement significatif de la complexité d'utilisation ; voir, à titre d'exemple, les dernières versions des logiciels de bureautique commercialisés. Ces évolutions rendent encore plus cruciale la nécessité d'aides en ligne qui permettent à l'utilisateur, novice ou occasionnel non informaticien, d'utiliser les logiciels grand public de manière efficace, et d'acquérir la maîtrise de leur utilisation, sans avoir à fournir un effort d'apprentissage spécifique important.

Or, si les logiciels courants actuels sont tous dotés d'aides en ligne évoluées qui mettent en œuvre, au moins en partie, les résultats des recherches effectuées dans les années 80 et au début des années 90, on constate que ces aides, en fait, sont ignorées ou rarement consultées. Les novices préfèrent recourir aux formations spécifiques florissantes pour acquérir les connaissances procédurales nécessaires à l'utilisation efficace d'un logiciel. Quant aux utilisateurs occasionnels, ils choisissent le plus souvent de faire appel à un collègue ou à un ami expérimenté pour résoudre un problème d'utilisation ponctuel ou sortir d'une situation de blocage. La raréfaction des recherches[2] sur l'aide à l'utilisation de logiciels constatée depuis le milieu des années 90 semble donc exprimer, dans ce contexte, un aveu d'impuissance à progresser dans la résolution des problèmes ergonomiques majeurs soulevés par l'aide en ligne, plutôt qu'un abandon de ce domaine de recherche motivé par la résolution satisfaisante de l'ensemble des problèmes de conception des aides en ligne.

La première partie de cet article situe la contribution des auteurs dans le débat de fond qui oppose depuis vingt ans deux points de vue antagonistes : les aides en ligne sont-elles un pis-aller, un simple remède aux défauts actuels de conception des logiciels et des interfaces utilisateur, ou bien, sont-elles une nécessité inhérente à l'utilisation d'un nouveau logiciel ? La seconde partie de cet article analyse les obstacles et les difficultés spécifiques auxquels se heurte la conception d'aides en ligne efficaces à l'intention du grand public, le critère majeur d'efficacité étant la consultation effective de l'aide par les utilisateurs, novices ou inexpérimentés, issus du grand public. Dans la dernière partie, les contributions potentielles de diverses approches à la résolution de ces problèmes spécifiques de conception sont analysées et discutées. Ces approches mettent en œuvre des paradigmes et techniques d'interaction récents, dont la multimodalité. La discussion fait intervenir, outre des arguments théoriques, des résultats empiriques et expérimentaux récents sur l'aide en ligne et l'interaction homme-machine multimodale. Ces résultats contribuent à valider les propositions de solutions présentées dans cette troisième partie.

---

[1] www.netcraft.com/
[2] constat confirmé au cours d'un entretien par Shneiderman.

# I. LES AIDES EN LIGNE : UN PIS-ALLER OU UNE NECESSITE ?

Historiquement et de façon schématique, on peut distinguer dans les travaux de recherche sur l'aide à l'utilisation des logiciels grand public, trois étapes principales :

- Une première conception dont l'apparition suit celles des ordinateurs personnels et de la manipulation directe (Shneiderman, 1983). L'aide en ligne y est vue comme un pis-aller, un mal nécessaire en l'absence d'interfaces utilisateur intuitives, « transparentes », un remède palliatif aux erreurs de conception. Ce point de vue a été défendu par de nombreux concepteurs de logiciels et d'interfaces utilisateur destinés au grand public.

- Dès le milieu des années 80, des études empiriques (voir la section I.2.1) portant sur l'analyse des comportements d'utilisateurs issus du grand public ont montré la nécessité d'assister l'utilisateur dans l'acquisition de la maîtrise de l'utilisation d'un logiciel nouveau pour lui.

- Ces résultats empiriques ont vraisemblablement stimulé les recherches sur l'aide en ligne. Des études (voir la section III.1) ont été entreprises pour faciliter l'accès des utilisateurs novices ou peu expérimentés aux informations d'aide et satisfaire leurs demandes/besoins d'assistance. Différentes approches exploitant des acquis en génie logiciel ou en intelligence artificielle ont été développées. Les travaux ont porté notamment sur :
    - l'organisation des informations d'aide sous la forme de bases de données, d'hypertextes ou d'hypermédias, et la mise au point de modes de consultation ou de navigation pertinents ;
    - et, plus récemment, la définition et l'implémentation de moyens d'interaction plus « naturels » ou « intelligents », tels que dialogues de consultation en langue naturelle, interfaces utilisateur adaptables ou adaptatives.

Les comportements du grand public, notamment ceux des utilisateurs novices et occasionnels, sont à l'origine des principales difficultés et obstacles auxquels sont confrontés les concepteurs de système d'aide en ligne à l'intention de cette population d'utilisateurs, et expliquent les échecs des différentes tentatives pour proposer à ces utilisateurs une assistance en ligne qui réponde à leurs besoins et attentes réels.

Les trois sections qui composent cette partie s'organisent comme suit. La première est une discussion des arguments initialement avancés par les chercheurs et concepteurs d'applications interactives destinées au grand public qui tiennent l'aide en ligne comme un palliatif au défaut de transparence des interfaces utilisateur existantes et aux erreurs de conception. La seconde présente brièvement des données empiriques et expérimentales qui démontrent la nécessité intrinsèque de l'assistance à l'utilisation d'un logiciel, nouveau pour l'utilisateur ou qu'il n'utilise qu'occasionnellement. La dernière résume les conclusions des deux précédentes.

## I.1. Interfaces transparentes : espoirs et déceptions

La manipulation directe s'efforce de combiner les possibilités offertes par les interfaces graphiques et la simplicité d'utilisation des systèmes à base de menus, en associant menus et icônes en référence à une métaphore de la vie courante (le bureau). De ce fait, elle représente une évolution radicale de l'interaction homme-machine qui, jusqu'à son apparition, imposait l'utilisation de langages de commande artificiels rébarbatifs dont la maîtrise était du ressort des seuls spécialistes. En offrant des moyens d'interaction simples et attractifs pour rendre l'informatique accessible au grand public, la manipulation directe a suscité des espoirs sans borne chez les concepteurs. En particulier, s'est développée la conviction qu'en améliorant encore davantage la transparence de l'interaction et la conception des interfaces utilisateur, on parviendrait à éliminer la nécessité de tout apprentissage initial, même réduit, à l'utilisation d'un nouveau logiciel (Mack et al., 1983 ; Mantei et Haskell, 1983 ; Carroll et Rosson, 1987). Dans sa comparaison de la manipulation directe avec les modes d'interaction antérieurs, Shneiderman défend ce point de vue lorsqu'il mentionne, parmi les principaux apports ergonomiques de ce nouveau mode d'interaction, la maîtrise rapide de l'utilisation d'un nouveau logiciel, la facilité d'apprentissage et d'assimilation de ses fonctions grâce à la simplicité de leur mise en œuvre (Shneiderman, 1987).

Les hypothèses suivantes qui ont présidé à la conception de la manipulation directe expliquent et justifient le point de vue de son créateur.
L'objectif majeur à l'origine de la manipulation directe était de fournir aux non informaticiens un mode d'interaction mieux adapté à leurs besoins et à leurs attentes que les langages de commande dont l'usage prévalait à l'époque. Cet objectif se fonde principalement sur l'hypothèse suivante. La syntaxe complexe et le vocabulaire arbitraire des langages de commande risquent de rebuter les utilisateurs novices ou occasionnels qui constituent la majorité du grand public, et donc de nuire à la diffusion de l'informatique dans la société. On peut supposer, en effet, que ces utilisateurs éprouveront des difficultés à apprendre et à assimiler des langages de commande artificiels dont la sémantique leur est étrangère en raison de leur ignorance des concepts informatiques exprimés et manipulés par ces langages. Un frein supplémentaire éventuel à l'utilisation de tels langages, pour les utilisateurs occasionnels, est leur manque de motivation pour acquérir les connaissances syntaxiques et lexicales nécessaires, en raison du coût élevé de cet apprentissage. Un tel investissement, en revanche, n'est pas de nature à décourager les utilisateurs fréquents expérimentés, dont les informaticiens. Un usage fréquent facilite l'assimilation et la mémorisation de ces connaissances arbitraires. En outre, l'efficacité de l'interaction étant l'une des préoccupations majeures de ces utilisateurs, la concision et la richesse expressive des langages de commande sont de nature à les séduire et à les motiver pour acquérir la maîtrise de leur utilisation. On trouvera dans (Shneiderman, 1987) une présentation détaillée des trois catégories d'utilisateurs considérées ici, novices, occasionnels, experts. Cette classification caractérise les utilisateurs de logiciels selon leurs compétences en informatique, leurs motivations et leur fréquence d'utilisation. Ces facteurs déterminent, selon l'auteur, leurs besoins et exigences spécifiques.
Une autre hypothèse a influencé profondément la conception de la manipulation directe. Elle peut se résumer comme suit. En offrant aux utilisateurs, grâce à une métaphore empruntée à l'expérience courante, la possibilité de manipuler des représentations graphiques pertinentes des objets de l'application considérée, on rend inutile tout apprentissage initial, en particulier syntaxique et lexical, de l'utilisation des logiciels qui opèrent sur ces objets. Cette hypothèse s'appuie sur les connaissances dont on dispose actuellement sur les capacités humaines d'adaptation qui permettent d'appliquer des savoirs et savoir-faire acquis dans une situation donnée à une situation nouvelle perçue comme semblable (Gick et Holyoak, 1983), que ce transfert mette en jeu des processus cognitifs de généralisation ou un raisonnement par analogie. Dans cette perspective, les utilisateurs novices non informaticiens seraient capables, à partir de leur expérience antérieure, de se familiariser avec un nouveau logiciel et d'acquérir la maîtrise de son utilisation par eux-mêmes, sans formation préalable ni effort d'apprentissage particulier, simplement en interagissant avec lui de manière naturelle et en explorant ses fonctionnalités à leur gré. En particulier, la métaphore « du bureau » est susceptible de faciliter le transfert de connaissances procédurales, puisqu'elle matérialise les informations électroniques, répertoires et fichiers, en les assimilant à des documents papier et des dossiers, objets d'usage courant aujourd'hui, au moins dans certaines sociétés. A noter en revanche que la manipulation à la souris des icônes graphiques 2D, utilisées pour représenter ces objets « papier » offre moins d'opportunités de transfert de savoir-faire antérieurs. Ce qui explique nombre des difficultés rencontrées par des utilisateurs inexpérimentés dans l'utilisation de la souris et des interfaces graphiques actuelles, par exemple lors de la découverte d'un traitement de texte (Capobianco et Carbonell, 2000).

Le point de vue décrit ci-dessus est encore partagé aujourd'hui par de nombreux chercheurs pour qui les défauts de conception (ou de réalisation) des logiciels et des interfaces utilisateur sont la principale raison d'être des systèmes d'aide en ligne (Chater, 1999). Dans la mesure où il persiste aujourd'hui, ce point de vue peut contribuer à expliquer pourquoi, depuis le milieu des années 90, la conception de nouveaux paradigmes d'interaction suscite davantage d'intérêt dans la communauté scientifique internationale que la définition et l'implémentation de nouvelles formes d'aide en ligne susceptibles d'accroître la flexibilité, l'efficacité et la robustesse des systèmes existants, prototypes de laboratoire et logiciels commercialisés.

La section suivante met en évidence les limites de ce point de vue et la nécessité d'une assistance à l'utilisateur dans la découverte d'un logiciel et la familiarisation avec son utilisation. L'argumentation se fonde principalement sur des données empiriques et expérimentales. Les études les plus anciennes, publiées peu après l'essor de la manipulation directe, sont centrées sur l'analyse des comportements des utilisateurs novices (section I.2.1). Une étude plus récente porte sur l'analyse des stratégies d'experts humains en situation d'assistance à l'utilisation d'un logiciel de traitement de texte que des novices découvrent (section I.2.2). La dernière section (I.2.3) présente des arguments théoriques qui contribuent à montrer la nécessité intrinsèque d'une assistance à la découverte et à la prise en main d'un logiciel inconnu, quelles que soient l'expérience antérieure de l'utilisateur et sa motivation.

## I.2. Nécessité de l'assistance à l'utilisateur novice

### I.2.1. Comportements d'utilisateurs novices

Quelques études empiriques réalisées dans les années 80 ont mis en évidence les limites de la manipulation directe en tant que mode d'interaction « transparent » et, indirectement, la nécessité d'une assistance à la découverte d'un nouveau logiciel, au moins pour les utilisateurs non informaticiens issus du grand public. Certaines d'entre elles (Carroll et McKendree, 1987 ; Fisher et al., 1985) ont montré que, contrairement aux attentes des concepteurs, les utilisateurs, en fait, n'exploitaient pas les possibilités d'exploration offertes par la manipulation directe, grâce à un feed-back visuel intuitif et immédiat, associé à la réversibilité des actions. (Streitz, 1988) a démontré en outre que la métaphore du bureau (*office*) pouvait induire les utilisateurs novices à se construire des représentations erronées des fonctionnalités du logiciel qu'ils découvrent. On trouvera dans (Beaudoin-Lafon, 1997) une discussion des limites des métaphores visuelles en interaction homme-machine. Il ressort des conclusions d'autres études que la majorité des utilisateurs issus du grand public ne parviennent pas à acquérir par eux-mêmes une connaissance suffisante du fonctionnement d'un logiciel nouveau pour leur permettre de réaliser les tâches qui motivent leur utilisation du logiciel. Par exemple, selon (Mack et al., 1983), les utilisateurs novices sont confrontés à de nombreuses difficultés d'ordre cognitif qu'ils ne peuvent surmonter seuls. Ils sont incapables, en particulier, de se construire une représentation exacte des capacités et du fonctionnement d'un logiciel nouveau. Des représentations erronées ou lacunaires sont à l'origine de nombreuses erreurs de nature sémantique que les novices ne peuvent ni corriger ni même détecter en l'absence des connaissances appropriées. Le rôle important joué par ces représentations mentales dans la prise en main d'un logiciel, bien qu'il ait été mis en évidence très tôt[3], n'est toujours pas pris en compte par les concepteurs d'interfaces utilisateur, peut-être en raison de la complexité de son influence sur l'interaction. La prévention et la correction automatiques des erreurs sémantiques sont d'ailleurs des objectifs extrêmement difficiles à atteindre. La diversité cognitive inter-individuelle humaine est telle qu'il est impossible aux concepteurs d'en tenir compte adéquatement. En effet, une des conditions déterminantes de l'efficacité de l'interaction avec un logiciel est la compatibilité entre deux représentations mentales des actions et tâches qu'il est possible d'effectuer sur les objets du domaine d'application : la vision du concepteur qui les automatise sous forme de fonctions logicielles et celle de l'utilisateur du logiciel. Or, le grand public constitue une population d'utilisateurs beaucoup trop hétérogène, en termes de capacités cognitives, de connaissances et d'expérience antérieures, pour qu'une même interface puisse satisfaire les besoins et attentes de l'ensemble d'entre eux. En outre, les concepteurs d'applications logicielles étant des experts [4], il y a de fortes chances pour que leur représentation du domaine d'application et de la tâche soit radicalement différente de celles d'utilisateurs novices ou occasionnels non informaticiens. Cette différence de représentation jointe à la grande diversité des capacités cognitives (apprentissage et adaptation notamment), des savoirs et savoir-faire des utilisateurs novices font qu'il est difficile de mettre en œuvre des stratégies d'interaction susceptibles de résoudre les problèmes d'utilisation qu'ils rencontrent. En pratique, les utilisateurs novices doivent donc nécessairement se familiariser avec la représentation conceptuelle du domaine d'application et de la tâche qui détermine la sémantique des fonctions logicielles proposées et qui est celle du concepteur. La réussite de cet « apprentissage » conditionne celle de la maîtrise de l'utilisation du logiciel. Ce point de vue conduit à considérer que l'un des rôles majeurs des aides en ligne est d'assister le novice dans l'acquisition de cette représentation et dans sa prise en compte pendant l'interaction avec le logiciel. Il contribue également à démontrer la nécessité, pour les utilisateurs novices ou occasionnels, de bénéficier d'aides en ligne qui leur facilitent cet apprentissage quelle que soit la transparence de l'interface utilisateur dont ils disposent. Il serait utile, en particulier, d'expliciter aux utilisateurs novices les relations entre :
– d'une part, leur représentation du domaine d'application sur lequel porte le logiciel considéré et des tâches/actions que celui-ci permet d'effectuer sur les objets de ce domaine, et
– d'autre part, celle du concepteur du logiciel.

---

[3] Voir la notion de *system image* dans (Sutcliffe et Old, 1987).

*I.2.2. Stratégies d'aide d'experts humains*

Les analyses que nous avons effectuées sur un corpus de dialogues entre des experts de l'utilisation d'un logiciel grand public et des utilisateurs novices confirment l'incapacité de ceux-ci à acquérir par eux-mêmes la maîtrise d'un logiciel grand public doté d'une interface graphique autorisant la manipulation directe, censée favoriser et faciliter l'exploration autonome des fonctions du logiciel (Capobianco et Carbonell, 2000, 2001). Les conclusions de ces analyses confirment, en particulier, que l'une des difficultés majeures des utilisateurs novices, qui constitue une source importante d'erreurs sémantiques, erreurs les plus difficiles à détecter et corriger, est leur incapacité à choisir, parmi les fonctions du logiciel, celles qui leur permettront de réaliser leurs intention et but courants.

Le corpus choisi comprend une quinzaine de dialogues expert-novice d'assistance à l'utilisation de Word. Les sujets, des étudiants en sciences humaines qui connaissaient superficiellement Windows (une demi-heure de manipulation) et qui n'avaient jamais utilisé Word avaient à réaliser vingt tâches de mise en page prédéfinies et classées par ordre de difficulté croissante. Deux experts étaient à leur disposition pour les aider à réaliser ces tâches. Leur rôle consistait à répondre à leurs questions ; ils ne devaient ni prendre le contrôle des échanges, ni intervenir spontanément de leur propre initiative, par exemple pour prévenir l'occurrence d'une erreur.
En vue d'étudier l'influence du contexte pragmatique du dialogue d'assistance sur le comportement et les stratégies d'aide des experts, trois situations ont été considérées et comparées. Dans l'une, expert et novice étaient assis dans la même pièce, côte à côte et face à l'écran du sujet. Dans les deux autres, en revanche, le contexte pragmatique était limité : sujet et expert étaient dans deux pièces différentes et communiquaient par interphone. Dans l'une de ces situations de communication à distance, les experts avaient accès aux affichages des sujets qui étaient reproduits sur leur écran grâce à un logiciel de copie d'écrans. Dans l'autre, ils disposaient uniquement de la liaison téléphonique, aucune information visuelle ne leur était fournie sur les interactions des novices avec le logiciel. Nous avons centré l'analyse sur les dialogues réalisés dans les deux dernières situations pour les raisons suivantes. D'une part, dans la situation où les experts avaient accès aux affichages du poste de travail des sujets (situation à contexte partagé limité), ils disposaient d'informations contextuelles similaires à celles auxquelles les systèmes d'aide peuvent accéder facilement. D'autre part, la comparaison entre les dialogues expert-novice réalisés dans cette situation et ceux recueillis dans la situation sans contexte visuel est susceptible de fournir des indications sur les informations contextuelles utilisées effectivement par les experts lors de leurs interventions d'assistance aux novices, et donc de faciliter l'élucidation, au moins partielle, de leurs stratégies d'aide [5].

---
[4] au sens de la classification proposée par Shneiderman.

| Actes de langage des experts humains | % |
|---|---|
| **Instruire** (condition B)<br>*E : Pour aligner l'adresse contre la marge de droite, il faut ...* | 40% |
| **Informer** (condition A)<br>*E : "Monsieur" n'est pas centré, devant il y a en fait un alinéa.* | 27% |
| **Evaluer** (condition A)<br>*N : Si je mets le curseur là, est-ce que je fais un retour à la ligne ?*<br>*E : Oui.* | 17% |
| **Demande d'information** (condition B)<br>*E : Est-ce que le curseur est devant "Monsieur" ?* | 8% |
| **Planification** (condition A)<br>*E : Maintenant, tu peux justifier le texte de la lettre.* | 8% |

Tableau 1 : Répartition des actes de langage des experts suivant une taxonomie *ad hoc* définie à l'issue d'une première analyse du corpus de dialogues. (Capobianco, 2002)

> Condition A : les affichages des sujets sont reproduits sur les écrans des experts.
> Condition B : les experts ne voient pas l'écran des sujets.
> ("E" signifie "Expert", et "N", "Novice").
> Les pourcentages ont été calculés sur l'ensemble des actes de langage des experts.

Les analyses ont été effectuées sur les transcriptions orthographiques des dialogues. Ces transcriptions incluent une brève description de l'évolution des affichages au cours de l'interaction. Tours de parole et affichages sont datés et présentés dans l'ordre de leur occurrence (Falzon et Karsenty, 1997). Les interventions orales des experts dans les deux situations à contexte limité (soit dix dialogues, cinq par situation) ont été décomposées en actes de langage dont le contenu informationnel a été caractérisé à l'aide des taxonomies existantes, notamment (Pilkington, 1992). La méthodologie est détaillée dans (Capobianco, 2002).

Les résultats sont résumés dans les tableaux 1 et 2 ; voir également (Capobianco et Carbonell, 2001). On constate, à la lecture de ces tableaux, que près de la moitié des actes de langage des experts (40%), et plus de la moitié des requêtes des novices (57%) portent sur des informations de nature procédurale. Ces résultats illustrent l'importance, dans les demandes d'aide exprimées par les novices, de leur besoin de clarifier la relation entre, d'une part, leurs représentations des tâches courantes de mise en page et, d'autre part, les fonctions et procédures du logiciel de traitement de texte qui permettent de les réaliser. Ils suggèrent également, dans l'hypothèse où les stratégies des experts humains sont optimales, qu'une des contributions majeures de l'aide à l'utilisation de logiciels est de faciliter à l'utilisateur la nécessaire évolution de sa représentation *a priori* des tâches réalisables grâce au logiciel. En effet, c'est dans le cadre de cette représentation initiale que l'utilisateur novice exprime les buts et intentions qui motivent son utilisation du logiciel, d'où l'importance cruciale de la compatibilité entre sa représentation du domaine d'application et celle du concepteur et, en cas d'incompatibilité, la nécessaire évolution de celle de l'utilisateur, puisque celle du concepteur, réifiée dans le logiciel, ne peut évoluer.

---

[5] Dans la situation où les experts n'ont pas accès aux affichages du poste de travail des sujets, les informations qu'ils sollicitent de la part des novices, notamment sur leur activité courante, présentent un intérêt particulier, car ce sont celles qu'ils utilisent pour mettre en œuvre leurs stratégies d'aide. Quant à la comparaison entre les deux situations où le contexte pragmatique est soit restreint, soit absent, elle est susceptible de fournir des informations utiles sur le rôle joué, dans les stratégies d'aide des experts, par les informations contextuelles présentes dans l'une des deux situations.

Ces résultats confirment les observations réalisées dès le début des années 80, en particulier le constat de l'incapacité des novices à acquérir seuls, par la pratique et l'exploration, la maîtrise de l'utilisation d'un logiciel. Ils enrichissent en outre ce constat, et contribuent à l'interprétation des comportements observés en identifiant l'une des causes principales à l'origine des difficultés des novices. Ils permettent enfin de progresser dans le débat sur la nécessité, intrinsèque ou non, d'aides en ligne à l'utilisation des logiciels grand public.

| **Actes de langage des utilisateurs novices** | **%** |
|---|---|
| **Comment ?** (condition A) | 57% |
| *N : Comment je fais pour remettre le curseur au début du texte ?* | |
| **Requêtes** | (37%) |
|    − Demande de confirmation (condition A) | 24% |
|      *N : Si je mets le curseur ici et tape CR, c'est OK?* | |
|    − Demande d'évaluation (condition A) | 12% |
|      *N : Est-ce que mon curseur est bien placé?* | |
|    − Demande d'assistance dans la planification (condition A) | 1% |
|      *N : Est-ce que je peux la déplacer (l'adresse) maintenant ?* | |
| **Quoi ?** (condition A) | 3% |
| *N : A quoi sert cette icône?* | |
| **Pourquoi ?** (condition B) | 3% |
| *N : Pourquoi ça l'a effacé (un mot du texte) ?* | |

Tableau 2 : Répartition des demandes d'informations des novices en fonction d'une taxonomie inspirée de (Pilkington, 1992). (Capobianco, 2002)

        "Conditions A et B", "E", "N" : voir la légende du tableau 1.
        Les pourcentages ont été calculés sur l'ensemble des actes de langage des novices.

En effet, l'évolution de la représentation initiale du novice vers une représentation qui soit compatible avec celle qui a présidé à la conception du logiciel ne s'impose pas seulement en raison des différences de point de vue et de connaissances entre informaticiens et utilisateurs du grand public, et/ou des limites éventuelles des techniques informatiques actuelles. Cet effort d'adaptation de la part du novice est indispensable si l'on ne veut pas cantonner l'informatisation d'une activité humaine à la reproduction de cette activité à l'identique. Si l'on vise une automatisation qui ne se réduise pas à un simple mimétisme de l'activité humaine correspondante, mais tente d'exploiter pleinement les capacités informatiques, très différentes des compétences humaines, on peut raisonnablement supposer que la plupart des utilisateurs novices ne parviendront pas, et même répugneront, à fournir l'effort nécessaire pour acquérir par eux-mêmes une représentation pertinente de la sémantique des fonctions des logiciels développés selon cette approche. En effet, si le fonctionnement d'un logiciel se démarque de la pratique « manuelle » de l'activité considérée et présente une réelle originalité par rapport à elle, le transfert de savoirs et savoir-faire antérieurs ne peut être mis à profit pour faciliter l'acquisition de la maîtrise de l'utilisation du logiciel ; seule une expertise en informatique est transférable dans ce contexte. L'utilisateur novice se trouve dans une situation analogue à celle du concepteur : pour parvenir à maîtriser l'utilisation du logiciel, il doit adopter une démarche créatrice et redécouvrir par lui-même les innovations que le logiciel implémente, alors qu'il ne dispose ni des compétences informatiques ni de la motivation nécessaires pour appliquer cette démarche. Dans cette perspective, les aides à l'utilisation de logiciels apparaissent donc comme une nécessité intrinsèque plutôt que comme un pis-aller, un mal nécessaire pour remédier aux défauts de conception des logiciels actuels et/ou au manque de transparence de leurs interfaces utilisateur.

La théorie écologique de la perception proposée par Gibson dans (Gibson, 1966) contribue également à démontrer les limites des approches de conception logicielle qui voient dans la mise en œuvre du concept de transparence la solution des problèmes d'utilisation auxquels sont confrontés les utilisateurs novices ou occasionnels. Les arguments que cette théorie fournit à l'encontre des espoirs suscités en interaction homme-machine par le concept de transparence sont présentés dans la section suivante.

### *I.2.3. Théorie écologique de la perception et transparence*

Les travaux de Gibson sur la perception ont été largement diffusés parmi les chercheurs en ergonomie des logiciels et ingénierie des interfaces homme-machine. Leur popularisation a donné lieu à de nombreuses définitions du concept d'« *affordance* » central dans la théorie complexe de Gibson. Parmi ces définitions, celle de Norman est brève et claire, au prix peut-être d'un certain réductionnisme :

> « … le terme *affordance* fait référence aux propriétés réelles perçues d'un objet, à savoir, essentiellement, les propriétés fondamentales qui déterminent les possibilités d'utilisation de l'objet. » [6] (Norman, 1988)

Or, les systèmes informatiques ne peuvent reproduire, avec des performances similaires, certaines activités humaines, par exemple celles qui relèvent de défis scientifiques majeurs en intelligence artificielle : perception et interprétation visuelles, compréhension de la parole, décision et raisonnement, mouvements et gestes complexes, etc. En outre, les environnements virtuels ne simulent que très imparfaitement le monde physique dont les propriétés d'*affordance* sont encore mal connues. Ces différences représentent des obstacles difficilement surmontables à l'implémentation du concept d'*affordance* en interaction homme-machine. Dans le cas des interfaces graphiques et des applications de bureautique, la métaphore du bureau (*office*) représente un moyen de vaincre ces obstacles, mais en partie seulement. En effet, dans ce contexte pourtant favorable, la métaphore est d'un apport plutôt limité, puisqu'il est nécessaire de lui associer un système de menus et de nombreuses boîtes de dialogue qui font appel aux ressources expressives de la langue naturelle. Plus généralement, les portées sémantique et pragmatique d'une métaphore ou d'une simulation sont, par définition, limitées, au même titre que l'analogie et la simulation diffèrent de l'identité et du clonage. La diversité des capacités cognitives, des connaissances et des savoir-faire antérieurs des utilisateurs potentiels représente un obstacle supplémentaire, majeur également. Par conséquent, la mise en œuvre du concept

---

[6] "… the term *affordance* refers to the perceived and actual properties of the thing, primarily those fundamental properties that determine just how the thing could be used."

d'*affordance* en interaction homme-machine dans le but d'offrir à l'utilisateur une interaction naturelle, transparente, semble difficile à réaliser. Les apports effectifs de ce concept à la conception des interfaces utilisateur sont limités, comme le soulignent les auteurs de (Weiser et Seely, 1995) qui estiment que le concept d'*affordance* malgré sa richesse ne rend compte que superficiellement de la relation entre un objet du monde réel et les intentions, les perceptions et les actions humaines qu'il a le potentiel de susciter [7].

### I.3. Conclusions

Pour conclure, les premières études empiriques ou expérimentales des comportements d'utilisateurs novices non informaticiens lors de la prise en main d'un logiciel grand public montrent que, malgré les possibilités d'exploration offertes par la manipulation directe et le caractère intuitif de ce mode d'interaction, ces utilisateurs sont incapables d'acquérir par eux-mêmes la maîtrise de l'utilisation d'un logiciel nouveau pour eux. Des données empiriques récentes confirment les conclusions de ces premières études et mettent en évidence une des causes majeures de l'incapacité des novices issus du grand public à maîtriser l'utilisation d'un nouveau logiciel par la seule pratique de l'interaction avec celui-ci, à savoir : l'incompatibilité entre, d'une part, leur représentation *a priori* des capacités et du fonctionnement du logiciel, élaborée à partir de leur expérience et connaissances antérieures, et d'autre part, les fonctionnalités effectives du logiciel que le concepteur informaticien a définies à partir de sa propre représentation du domaine d'application considéré et des opérations réalisables sur les objets de ce domaine. La discussion de ces résultats conduit à considérer l'aide à l'utilisation de logiciels comme une nécessité intrinsèque dont l'un des rôles principaux est de faciliter aux utilisateurs novices non informaticiens l'évolution nécessaire de leur représentation initiale, partielle et/ou erronée des fonctionnalités du logiciel vers une représentation compatible avec le fonctionnement effectif de celui-ci, évolution ou adaptation qu'ils ne peuvent effectuer seuls dans la mesure où les connaissances sémantiques qu'elle met en jeu leur font défaut.

En concluant à la nécessité d'une aide pour résoudre les difficultés de nature sémantique inhérentes à l'utilisation de tout nouveau logiciel plutôt que pour remédier aux défauts de conception du logiciel et/ou au manque de transparence de son interface utilisateur, on met implicitement en cause la possibilité de réaliser des interfaces utilisateur totalement transparentes. La discussion, dans la dernière section de cette partie, des possibilités d'application du concept d'*affordance* en interaction homme-machine confirme cette conclusion en mettant en évidence la difficulté et l'intérêt pratique limité de la mise en œuvre de ce concept. La transparence de l'interaction apparaît donc comme un objectif de conception irréaliste, qu'il est illusoire d'espérer atteindre grâce au progrès scientifique, technique ou technologique.

La partie suivante est consacrée à la présentation et à l'analyse des problèmes spécifiques que pose la conception de systèmes d'aide en ligne efficaces, donc effectivement utilisés.

## II. CONCEPTION D'AIDES EN LIGNE : DEFIS ET VERROUS SPECIFIQUES

La conception des systèmes d'aide à l'utilisation de logiciels soulève des problèmes spécifiques supplémentaires par rapport à ceux que doivent résoudre les concepteurs de logiciels interactifs classiques pour satisfaire les besoins et les attentes de la population d'utilisateurs ciblée.

---

[7] "An *affordance* is a relationship between an object in the world and the intentions, perceptions, and capabilities of a person … The idea of *affordance*, powerful as it is, tends to describe the surface of a design. For us the term '*affordance*' does not reach far enough into the periphery where a design must be attuned to but not attended to."

A l'origine de ces difficultés, trois causes principales :

- La nature même de la tâche, à savoir, fournir aux utilisateurs inexpérimentés les informations dont ils ont réellement besoin au moment exact où ils en ont besoin.
- La situation d'interaction, où l'utilisateur novice est simultanément impliqué dans deux tâches différentes : l'interaction avec le logiciel d'une part, et la consultation de l'aide d'autre part.
- Les comportements et les attitudes des utilisateurs non informaticiens qui constituent en majorité le grand public.

Ces spécificités et leur incidence sur la conception des systèmes d'aide sont détaillées dans les trois sections suivantes.

## II.1. Pertinence des informations et des interventions d'aide

Fournir aux utilisateurs novices les informations procédurales et les méta-informations dont ils ont besoin pour réaliser les tâches qui motivent leur recours à un logiciel au moment où elles leur sont nécessaires, est un défi spécifique auquel les concepteurs de systèmes d'aide sont confrontés.

Méta-information et méta-communication font référence ici aux informations et échanges d'informations définis dans le modèle conceptuel des fonctionnalités du logiciel. Ces concepts ont été proposés et leurs implications en interaction homme-machine discutées au cours des années 80 (Tauber, 1986 ; Waern, 1989). Malheureusement, ils n'ont exercé aucune influence sensible sur la conception des systèmes d'aide en raison, semble-t-il, de l'ignorance ou de la méconnaissance des concepteurs. Pourtant, ils sont susceptibles de jouer un rôle important dans la définition du contenu informationnel des messages d'aide et dans leur mise en correspondance temporelle avec les besoins effectifs des novices.

En effet, les études empiriques et expérimentales du comportement des utilisateurs novices non informaticiens ont montré que ces utilisateurs avaient des difficultés à se construire une représentation suffisamment fidèle des fonctionnalités d'un logiciel inconnu d'eux pour qu'elle leur permette d'acquérir par eux-mêmes la maîtrise de son utilisation (voir les sections I.2.1 et I.2.2). En outre, la consultation du système d'aide qu'ils découvrent en même temps que le logiciel représente pour eux une source de difficultés supplémentaires (Kearsley, 1988 ; Sellen et Nicol, 1990). Ils sont incapables, seuls, de « trouver » l'information procédurale spécifique dont ils ont besoin pour achever la tâche courante. Ils éprouvent des difficultés à « comprendre » des messages d'aide génériques, formulés dans les termes même du modèle conceptuel des fonctions du logiciel, souvent parce qu'ils ne disposent pas des méta-connaissances informatiques nécessaires à leur compréhension. Pour les aider à surmonter ces difficultés, (Waern, 1985) propose une assistance comprenant, outre une présentation initiale des fonctionnalités du logiciel, un traitement des échecs de leurs actions et l'accès à des informations appropriées sur le contexte de leurs interactions, notamment sur leurs actions antérieures (historique des actions). Concernant les difficultés qu'ils rencontrent dans la compréhension des méta-informations d'aide, de nombreux guides de conception proposent des recommandations en vue d'assurer l'intelligibilité des messages d'aide. La plupart de ces recommandations, qui relèvent du bon sens, portent sur l'élaboration des messages textuels ; voir (Carroll et al., 1987), par exemple.

## II.2. Situation et contexte d'interaction

Une caractéristique spécifique de la situation d'interaction avec une aide en ligne est la contrainte, pour l'utilisateur, d'interrompre son activité principale, l'interaction avec le logiciel, chaque fois qu'il consulte l'aide. Ces allers-retours entre l'application et le système d'aide sont une source importante de difficultés, surtout pour les utilisateurs novices (Kearsley, 1988 ; Sellen et Nicol, 1990). L'interruption de l'activité principale, bien que décidée par l'utilisateur, est de nature à perturber cette activité et nuire à son efficacité. Il peut la ressentir négativement, comme une contrainte, dans la mesure où elle lui est souvent dictée par l'impossibilité de poursuivre l'interaction avec le logiciel, par exemple en raison d'une erreur d'utilisation bloquante. Cette contrainte peut le conduire à limiter ses recours au système d'aide, et contribuer à expliquer la faible utilisation des aides en ligne.

Les interruptions de l'activité principale liées à la consultation de l'aide peuvent donc s'avérer aussi pénalisantes que les alertes et la notification d'événements dans certains environnements. Or, les interruptions occasionnées par la notification d'événements extérieurs à l'application courante présentent un caractère intrusif marqué dont le coût pour l'utilisateur préoccupe les concepteurs. La gestion optimale de ces interruptions suscite un intérêt croissant de la part des chercheurs en interaction homme-machine (Horvitz et Apacible, 2003 ; McCrickard et al, 2003).

## II.3. Comportements et attitudes des novices

L'obstacle spécifique majeur à l'utilisation effective des aides en ligne, donc à leur efficacité, a été mis en évidence dès les premières observations de l'utilisation de la manipulation directe et des logiciels de bureautique par des utilisateurs non informaticiens relevant du grand public. Il s'agit du « paradoxe de motivation » qui, selon (Carroll et Rosson, 1987), caractérise les comportements et les attitudes de cette catégorie d'utilisateurs lors de la découverte et de la prise en main d'un nouveau logiciel.

Carroll et Rosson concluent, à partir de l'analyse de données empiriques variées, que les utilisateurs issus du grand public, les novices en particulier, lorsqu'ils prennent en main un nouveau logiciel, ne tentent pas d'explorer ses fonctionnalités. Ils ne s'appliquent pas davantage à maîtriser son utilisation. Dès les premières interactions, ils s'investissent essentiellement dans la réalisation des objectifs et des tâches qui motivent leur recours au logiciel. L'efficacité de leur interaction avec le logiciel, son optimisation, n'entrent pas dans leurs préoccupations majeures. Ils ignorent les incitations à l'exploration qu'offre la manipulation directe et, plus généralement, les possibilités d'apprentissage autonome, par essai-erreur, que favorisent les interfaces où les actions de l'utilisateur sont réversibles. Selon ces chercheurs, le paradoxe de motivation expliquerait pourquoi les utilisateurs qui composent le grand public, notamment les novices, consultent rarement ou pas du tout les aides en ligne. Ce paradoxe constitue l'obstacle principal que les concepteurs ont à surmonter pour réaliser des systèmes d'aide qui soient effectivement utilisés. L'utilisation effective d'une aide en ligne est un objectif dont la satisfaction doit être prise en compte dans les critères classiques d'évaluation de la qualité ergonomique de celui-ci.

En dépit de l'évolution des aides en ligne depuis les travaux de Carroll et Rosson, les systèmes d'aide à l'utilisation de logiciels grand public actuels ne sont pas davantage utilisés que ceux des années 80, et le paradoxe de motivation est toujours pertinent pour rendre compte du comportement et de l'attitude du grand public à l'égard des systèmes d'aide. Pourtant, des efforts considérables et de nombreuses tentatives ont été effectués pour faciliter la consultation des aides en ligne et développer leur utilisation effective. En particulier, on a substitué aux manuels classiques organisés hiérarchiquement :
− des bases de données relationnelles, plus faciles à consulter car elles autorisent un accès associatif simple, par mots clés, aux informations d'aide (Fisher, Lemke et Schwab, 1985) ;
− ou des hypertextes (Cohill et Williges, 1985 ; Moore et Swartout, 1990) et, plus récemment, des hypermédias (Palmiter et Elkerton, 1991 ; Harrison, 1995) qui mettent en œuvre un accès associatif plus intuitif que les bases de données classiques ; en effet, on peut considérer les liens inclus dans un document hypertexte ou hypermédia comme des mots-clés contextuels, le contexte dans lequel ils apparaissent permettant de lever facilement les ambiguïtés dues à la polysémie du vocabulaire des langues naturelles.

Les systèmes d'aide actuels des logiciels grand public commercialisés mettent en œuvre ces différentes techniques d'accès, souvent en parallèle, de façon à laisser le choix du mode de consultation à l'utilisateur ; mais cette évolution semble sans effet sur les pratiques de la plupart de leurs utilisateurs.

Carroll et ses collègues ont proposé une solution originale pour résoudre les problèmes spécifiques de conception créés par le paradoxe de motivation. Malheureusement, cette solution n'a pas été expérimentée jusqu'à présent, vraisemblablement en raison des difficultés que présente sa mise en œuvre. Elle consiste à donner aux novices les moyens d'accomplir, dès la prise en main du logiciel, les tâches permettant la réalisation de leurs intentions et de leurs buts, en leur évitant tout apprentissage initial systématique de l'utilisation du logiciel (Carroll et al., 1987). Cette approche implique la mise en œuvre d'une assistance à l'activité, voire même d'une véritable coopération homme-machine, difficile à réaliser, surtout avec des utilisateurs novices, car elle nécessite la connaissance des but et intention courants de l'utilisateur. Or, les intentions des utilisateurs inexpérimentés sont souvent masquées par les nombreuses erreurs d'utilisation (syntaxiques ou sémantiques) qu'ils commettent.

Dans la partie suivante sont évoquées et discutées plusieurs approches envisageables pour résoudre les trois problèmes spécifiques que soulève la conception des systèmes d'aide en ligne.

## III. VERS DES AIDES EN LIGNE EFFICACES

Il ressort de la partie précédente que, pour réaliser des aides en ligne efficaces, effectivement consultées, les concepteurs doivent viser les objectifs spécifiques suivants :

- Faciliter au novice l'acquisition des connaissances et savoir-faire nécessaires pour utiliser efficacement un nouveau logiciel, sans que cet apprentissage ne perturbe son interaction avec le logiciel, centrée sur la réalisation des tâches qui motivent son utilisation du logiciel.
- Prendre en compte la diversité cognitive inter-individuelle, de façon à permettre à l'ensemble des utilisateurs non informaticiens d'acquérir facilement, rapidement et de façon durable ces connaissances et savoir-faire en exploitant intensivement l'interactivité (Carrol et Mack, 1992). Ce qui implique, notamment :
    - la prise en compte de la grande diversité des connaissances initiales de ces utilisateurs,
    - une assistance effective à l'acquisition, grâce à l'interaction avec le logiciel, d'une représentation exacte des fonctionnalités de celui-ci.
- Favoriser le développement de processus d'apprentissage par l'action ou *learning by doing* (Lewis et Williams, 1994) et d'apprentissage incident ou *incidental learning* (Ross-Gordon et Dowling, 1995), en centrant l'aide sur l'assistance à la réalisation des buts et intentions des novices. On peut espérer, en procédant ainsi, parvenir à surmonter le paradoxe de motivation (voir section II.3).

A noter que le paradoxe de motivation, qui caractérise l'attitude de la majorité des utilisateurs novices vis-à-vis de l'aide en ligne, fait de l'acquisition interactive de la maîtrise de l'utilisation d'un logiciel une situation spécifique et nouvelle. Ainsi, les résultats des nombreux travaux de recherche sur les environnements logiciels classiques d'assistance à l'apprentissage de connaissances ou de savoir-faire (voir Norris et al, 2002, pour une synthèse de ces recherches), de même que les résultats de travaux plus récents, consacrés aux « tuteurs intelligents » (par exemple, Cherniavsky et Soloway, 2002) ou à l'analyse des stratégies didactiques des enseignants en formation initiale ou continue (Rogalski, 2003), ne peuvent s'appliquer à cette situation. Il en va de même pour les études empiriques ou expérimentales sur l'apprentissage de l'informatique, notamment la programmation (Corritore et Wiedenbeck, 2001, par exemple) et

la conception (Détienne et al., 2005), car ces études analysent et modélisent les comportements d'informaticiens professionnels plutôt que ceux du grand public, dans des contextes d'apprentissage explicite plutôt qu'implicite.

Les sections qui suivent sont consacrées à une discussion de la contribution potentielle des principales approches qu'il est raisonnable d'envisager actuellement pour surmonter les obstacles spécifiques qui limitent l'efficacité des aides en ligne :
− l'adaptation statique ou dynamique de l'aide au profil de l'utilisateur courant ;
− la mise en œuvre des principes qui définissent l'approche dite de conception universelle (*universal design*) ;
− la prise en compte du contexte de l'interaction dans la conception et la présentation des messages d'aide, ou aide contextuelle ;
− l'apport potentiel de la multimodalité à l'efficacité de l'aide en ligne.

### III.1. Adaptation statique et dynamique

Le concepteur a le choix entre deux familles principales de techniques pour adapter les messages d'aide et, plus généralement les réactions du système, aux capacités, connaissances et préférences de l'utilisateur courant :

- L'adaptation statique (*adaptability*), qui utilise un modèle de l'utilisateur prédéfini. Les modèles statiques utilisés se fondent le plus souvent sur une classification des utilisateurs potentiels du logiciel considéré. Chaque classe ou stéréotype caractérise, par rapport aux critères choisis, les comportements similaires des utilisateurs qui la composent. L'affectation d'une classe à l'utilisateur courant peut être à la charge, soit de l'utilisateur (*customization*), soit du système. La classification grossière des utilisateurs de logiciels proposée par Shneiderman est fréquemment utilisée pour offrir à l'utilisateur des possibilités d'adaptation statique. C'est le cas, par exemple, des logiciels grand public courants qui proposent à la fois des menus simplifiés aux novices et aux utilisateurs occasionnels, et des raccourcis clavier aux utilisateurs experts. On trouvera dans (Sutcliffe et Old, 1987) une définition du concept de stéréotype.

- L'adaptation dynamique (*adaptivity*) utilise un modèle adaptatif de l'utilisateur, c'est-à-dire un modèle capable d'évoluer au cours de l'interaction et, en particulier, de s'adapter à l'évolution des savoirs, savoir-faire et attitudes de l'utilisateur courant au fur et à mesure de sa pratique de l'interaction et du progrès de sa découverte du logiciel. La mise à jour de ces modèles utilise essentiellement les indices qu'il est possible d'extraire de la trace des interactions (Finck et al., 1997 ; Langley, 1999) et, éventuellement, des informations spécifiques fournies par l'utilisateur en réponse à des requêtes du système.
Jameson (2003) a effectué une synthèse remarquable des recherches menées dans le cadre de cette thématique. L'auteur met en évidence, dans des domaines d'application variés, les possibilités offertes aux concepteurs de systèmes interactifs par les techniques d'adaptation dynamique ; il montre la puissance de ces techniques et précise leurs limites en majorité d'ordre ergonomique.

Un ensemble réduit de stéréotypes ou de classes d'utilisateurs ne peut manifestement pas représenter de manière adéquate la très grande diversité des profils cognitifs des utilisateurs formant le grand public. Seuls, les modèles dynamiques de l'utilisateur possèdent la flexibilité nécessaire pour rendre compte d'une diversité inter-individuelle aussi importante. En outre, les systèmes d'aide doivent, en raison même de leur finalité, être capables de s'adapter à l'évolution des comportements et des compétences qui accompagne, pour un même utilisateur, l'acquisition de la maîtrise du logiciel sur lequel porte l'assistance. L'adaptation à cette variabilité intra-individuelle, inhérente aux situations de prise en main d'un logiciel, nécessite la capacité d'identifier, pendant l'interaction, l'état courant des compétences de l'utilisateur.

Les systèmes adaptatifs sont capables, potentiellement, de réaliser une telle adaptation. Cependant, la conception et la réalisation de systèmes dont le comportement s'adapte effectivement à l'évolution des compétences de l'utilisateur courant est un objectif beaucoup plus difficile à atteindre que la prise en compte dynamique des variations de ses intérêts et de ses préférences. La difficulté principale est de parvenir à inférer la nature et l'évolution exactes des compétences de l'utilisateur novice à partir de l'analyse de la trace de ses interactions. Cette difficulté peut contribuer à expliquer pourquoi la majorité des tentatives d'application du concept d'adaptation dynamique ont porté jusqu'à présent sur la réalisation d'interfaces adaptatives de consultation du Web, dont l'objectif principal est de filtrer la masse des informations accessibles en fonction des intérêts des utilisateurs ; voir (Lieberman et Maulsby, 1996), l'une des premières mises en œuvre de cette forme de personnalisation. Plus proche des préoccupations des concepteurs d'aides à l'utilisation de logiciels, le prototype décrit dans (Espinoza et Höök, 1996) effectue un filtrage personnalisé du contenu d'un manuel en ligne (500 pages HTML) d'assistance à l'utilisation d'une méthode de développement logiciel.

Quelques prototypes adaptatifs d'aide en ligne ont certes été développés au début des années 90 (Brajnik et Tasso, 1994 ; Paiva et Self, 1994). Ils utilisent des techniques sophistiquées d'intelligence artificielle, telles que raisonnement sur des données incertaines, vérification et maintien de la cohérence d'un ensemble d'informations (*truth maintenance systems*), pour inférer, à partir des informations fournies par l'historique des interactions de l'utilisateur novice avec le logiciel qu'il découvre et de ses requêtes au système d'aide, ses progrès dans l'acquisition des connaissances procédurales et déclaratives nécessaires pour maîtriser l'utilisation d'un nouveau logiciel. Cependant, ces prototypes sont trop complexes pour que les utilisateurs, en particulier les novices, puissent se construire, en interagissant avec eux, une représentation exacte de leurs capacités et de leur fonctionnement. Sans cette connaissance, les réactions du système risquent de leur paraître imprévisibles et incompréhensibles, ce qui va à l'encontre de l'un des principaux critères classiques utilisés en ergonomie des logiciels (Shneiderman, 1987). Le manque de fiabilité est une autre faiblesse majeure de ces prototypes si l'on applique les critères ergonomiques classiques.

Une approche originale mérite d'être signalée (Höök, 1995). Elle consiste à structurer les informations d'aide en fonction des tâches réalisables dans le domaine d'application considéré, puis à prévoir, pour chaque tâche élémentaire, plusieurs ensembles d'informations d'aide atomiques correspondant aux différents profils des utilisateurs de la population ciblée (connaissances générales, expérience, familiarité avec le logiciel, entre autres). En combinant dynamiquement ces informations en fonction de la tâche, plus ou moins complexe, en cours de réalisation, on obtient plusieurs messages (un par profil utilisateur) parmi lesquels l'utilisateur choisit celui qui lui correspond. Pour que l'utilisateur soit à même de choisir le niveau d'information le mieux adapté à son profil, il faut que les différences entre les divers niveaux d'information proposés pour une même tâche soient clairement indiqués. On obtient ainsi un système d'aide hybride, adaptatif pour autant qu'il fournit des informations ajustées dynamiquement à la tâche courante, et adaptable dans la mesure où c'est l'utilisateur qui choisit le niveau d'information. Cette approche est séduisante *a priori*. Cependant, elle est difficile à mettre en œuvre car elle oblige à analyser et structurer un domaine d'informations dont la taille est fonction de celle du domaine d'application considéré, et à définir des profils utilisateurs détaillés et spécifiques (car liés au domaine d'application), pour être en mesure de concevoir des niveaux d'informations pertinents. L'entreprise risque d'être gigantesque dans le cas d'applications réelles, d'autant plus que l'acquisition des connaissances nécessaires ne peut être automatisée. En outre, la construction des ontologies est à refaire pour chaque nouveau domaine d'application abordé. Ceci explique probablement pourquoi cette approche ne s'est pas développée.

Pour conclure, l'application du concept d'adaptation statique à la conception des systèmes d'aide ne permet pas de résoudre les problèmes spécifiques évoqués dans la seconde partie de cet article. L'adaptation dynamique apparaît en revanche comme un concept approprié pour surmonter ces obstacles. Malheureusement, les techniques actuelles ne permettent pas de la mettre en œuvre de façon satisfaisante. Des recherches longues et difficiles sont à mener dans ce domaine avant d'envisager l'introduction sur le marché de systèmes d'aide en ligne adaptatifs. Il n'est même pas certain que le concept d'adaptation dynamique recueille l'adhésion de la majorité des utilisateurs issus du grand public. L'accueil réservé des utilisateurs aux menus adaptatifs des récentes versions des logiciels Microsoft [8] est de nature à susciter une certaine inquiétude quant au succès de la mise en œuvre de ce concept en interaction homme-machine.

### III.2. Mise en œuvre du paradigme de conception universelle

L'application en interaction homme-machine du paradigme de conception universelle (*universal design*) tel qu'il est défini dans (Connell et al., 2001) apparaît comme une alternative séduisante aux techniques d'adaptation pour assurer l'utilisation effective des aides en ligne. Selon ses créateurs, le paradigme de conception universelle représente un concept et un ensemble d'expériences utiles pour promouvoir le développement d'une Société de l'Information planétaire qui soit accessible à tous (*universal access*). L'objectif est en effet de concevoir des produits et des environnements qui soient utilisables par tout le monde sans adaptation particulière ni aménagement spécifique [9]. Conçu initialement à l'intention des architectes, il n'est pas surprenant que ce paradigme de conception et les sept principes qui définissent sa mise en œuvre réduisent au minimum le rôle de l'adaptation, contrairement aux paradigmes de « conception centrée sur l'utilisateur » (*user centred design*) ou de « conception pour tous » (*design for all*) ; voir, pour une définition de ces approches, (Bevan et Curson, 1997) et (Bevan, 2001), respectivement.

On peut être tenté, *a priori*, d'appliquer l'approche dite de conception universelle à la réalisation des systèmes d'aide en ligne, dans la mesure où, malgré leur grande diversité, les utilisateurs qui découvrent un logiciel possèdent une caractéristique commune : leur ignorance des capacités, de l'utilisation et du fonctionnement de ce logiciel. Le principal avantage de cette approche est sa simplicité de mise en œuvre. Cette simplicité est propre à garantir la fiabilité et la robustesse du système d'aide. En outre, elle facilite aux utilisateurs la construction d'une représentation exacte des capacités et du fonctionnement du logiciel, nécessaire pour que ceux-ci soient en mesure de prédire les réactions du système. La mise en œuvre du paradigme de conception universelle apparaît donc comme une voie prometteuse pour surmonter l'un des obstacles spécifiques auxquels sont confrontés les concepteurs de systèmes d'aides en ligne, la diversité cognitive inter- et intra-individuelle des utilisateurs de ces systèmes. Cependant, elle ne semble pas en mesure de résoudre le dernier problème spécifique de conception évoqué dans la section II.3, celui qui résulte des comportements et attitudes des utilisateurs vis-à-vis de l'aide en ligne ou, pour reprendre l'expression utilisée dans (Carroll et Rosson, 1987), du paradoxe de motivation.

On peut toutefois espérer résoudre ce problème sans renoncer à l'application du paradigme de conception universelle, et pouvoir offrir aux utilisateurs novices les moyens de développer un apprentissage implicite [10] de l'utilisation d'un nouveau logiciel, grâce à la mise en œuvre du concept d'aide contextuelle. Un système d'aide en ligne contextuelle s'efforce, et doit être capable, de familiariser l'utilisateur novice avec l'utilisation d'un nouveau logiciel tout en l'assistant dans la réalisation interactive des tâches qui motivent son recours au logiciel. Les premières tentatives de mise en œuvre du concept d'aide en ligne contextuelle datent des années 80 (Dzida et al., 1987). Mais aucun des rares prototypes dont la description a été publiée n'a fait l'objet d'une véritable évaluation ergonomique. Il est donc difficile, en l'absence d'une telle évaluation, de déterminer l'efficacité réelle de cette approche pour résoudre le problème de conception résultant du paradoxe de motivation.

---

[8] Voir, par exemple, les versions de Pack Office postérieures à 1998.
[9] Objectifs visés par ce paradigme (Story, 1988, page 4) : "… the design of products and environments to be usable by all people, to the greatest extent possible, without the need for adaptation or specialized design."
[10] C'est-à-dire un apprentissage incident ou fondé sur l'action et la pratique ; voir l'introduction de la section III.

La section suivante contribue à combler cette lacune en présentant une évaluation ergonomique récente d'un système d'aide en ligne (simulé) qui applique une stratégie d'aide contextuelle pour assister les utilisateurs novices d'un logiciel grand public.

### III.3. Aides en ligne contextuelles

#### *III.3.1. Informations contextuelles utilisées*

La mise en œuvre de stratégies d'aide contextuelles implique, outre la connaissance de l'état courant du logiciel, celle du but (donc de la tâche) et de l'intention (donc de l'action) que l'utilisateur courant vise à réaliser. C'est du moins l'une des conclusions majeures de notre analyse des dialogues expert-novice d'assistance à l'utilisation de Word (voir la section I.2.2). Ce sont ces informations que les experts humains utilisent implicitement pour définir le contenu informationnel de leurs réponses aux requêtes des novices (Capobianco, 2002). Le tableau 3 précise les différents types d'informations contextuelles qu'ils mettent en oeuvre pour définir le contenu informationnel de leurs réponses aux demandes d'informations procédurales des utilisateurs novices.

Les experts semblent appliquer effectivement une stratégie d'aide contextuelle, au sens où nous avons défini cette forme d'aide en ligne dans la section précédente. En effet, d'après les données présentées dans le tableau 3, le choix du contenu de leurs apports d'informations procédurales est influencé en priorité par l'état de réalisation de la tâche en cours (71% de leurs interventions) et, à un degré moindre, par l'intention courante du novice (19%) et l'état de l'application (17%). En outre, 46% de leurs actes de langage comprennent des références explicites au contexte de l'interaction (action courante du novice, état de réalisation de la tâche en cours, état courant du système, notamment). Enfin, près des deux tiers d'entre eux visent à assister le novice dans son activité de mise en page puisqu'ils comprennent :

− 40% d'apports d'informations procédurales,
− 17% d'évaluations des effets des actions du novice sur l'interface utilisateur, et
− 8% de tentatives d'assistance à la planification de son activité.

En revanche, moins d'un tiers d'entre eux (27%) sont des apports d'information générale sur le fonctionnement du logiciel ou la sémantique de ses fonctions, à la différence des didacticiels et des tutoriels d'experts humains qui tendent à privilégier ce type d'informations.

| Action courante | Etat courant du système | Etat de réalisation de la tâche courante | Historique (dialogue et interactions) | Intention courante | Autre |
|---|---|---|---|---|---|
| 3% | 17% | 71% | 5% | 19% | 9% |

Tableau 3 : Nature des informations contextuelles mises en œuvre par les experts humains. (Capobianco, 2002)

> Les pourcentages ont été calculés sur les actes de langage des experts de type apport d'informations procédurales (soit 40% de l'ensemble de leurs actes de langage). Certains actes de langage font intervenir plusieurs types d'informations contextuelles, ce qui explique les résultats.

Cependant, la mise en œuvre de la stratégie d'aide contextuelle des experts humains, stratégie que l'on est tenté de considérer comme idéale, présente des difficultés considérables. C'est pourquoi, vraisemblablement, les prototypes dotés de stratégies évoluées d'exploitation du contexte de l'interaction, comme ceux décrits dans (Carenini et Moore, 1993) et (Quast, 1993) sont rares et leur usage reste limité.

### *III.3.2. Difficultés de réalisation*

La difficulté majeure qui s'oppose à la réalisation d'aides contextuelles réside dans l'identification des buts et intentions de l'utilisateur à l'origine de ses actions sur le logiciel, et le suivi de leur évolution au cours de l'interaction. Cette évolution est encore plus difficile à cerner que celle des progrès de l'utilisateur novice dans la maîtrise de l'utilisation du logiciel.

La plupart des approches tentent de détecter des régularités ou motifs (*patterns*) dans la trace des interactions. Ces régularités sont ensuite interprétées en tant qu'indices à partir desquels sont inférés les intentions et buts courants de l'utilisateur. La recherche de motifs met en œuvre, le plus souvent, des techniques d'apprentissage et de reconnaissance statistiques (Horvitz et al., 1998 ; Jameson et al., 2000). Or, pour être en mesure d'assister l'utilisateur novice dans l'utilisation d'un logiciel, il faut connaître non seulement son intention courante, mais aussi la tâche et le but global dans lesquels elle s'inscrit. Par exemple, dans le cas de la réalisation d'une procédure complexe, il faut être capable d'identifier le plus tôt possible le but que l'utilisateur cherche à réaliser en exécutant la procédure. L'identification précoce du but global de l'utilisateur novice et même la détection de son intention courante sont toujours difficiles, en raison des nombreuses erreurs d'utilisation que celui-ci commet par ignorance des capacités et du fonctionnement du logiciel. Elle est même impossible, dans le cas des logiciels grand public courants, quel que soit le niveau de compétence de l'utilisateur, car ceux-ci offrent un ensemble réduit de fonctions élémentaires qui doivent être combinées entre elles pour réaliser les nombreuses tâches, plus ou moins complexes, relevant du domaine d'application couvert par le logiciel. Par conséquent, la première action qu'entreprend l'utilisateur pour réaliser un but est souvent ambiguë, car elle peut être le point de départ de la réalisation de tâches très diverses. Par exemple, il est impossible d'inférer, à partir de la sélection d'un objet graphique, l'opération que l'utilisateur envisage d'effectuer sur lui.

La seule solution possible pour connaître l'objectif et l'intention courante du novice, lorsque la trace des interactions ne permet pas de les inférer, est d'inciter celui-ci à les exprimer avant d'entreprendre leur réalisation ou dès les premiers stades de celle-ci. Cette stratégie est celle qu'ont fréquemment adoptée les experts humains des dialogues d'assistance que nous avons analysés (voir l'étude présentée dans la section I.2.2), puisque 8% de leurs actes de langage visent à obtenir ce type d'information. La mise en œuvre de tels dialogues de clarification ne semble pas présenter de difficultés insurmontables, car ils sont généralement très courts, et leur structure très simple, de type question-réponse. Comme ils sont à l'initiative du système, leur contrôle est facile à assurer par celui-ci. Enfin, leur caractère très ciblé limite la complexité des énoncés à interpréter et à générer. La seule véritable difficulté pourrait venir des visions différentes que le concepteur et les utilisateurs novices ont du domaine d'application et des tâches réalisables (voir la section I.2.1). Différentes solutions peuvent être envisagées pour surmonter cette difficulté. Une direction de recherche à explorer serait d'étudier comment doter le système d'aide de connaissances suffisantes sur les représentations erronées des fonctionnalités du logiciel les plus fréquentes parmi les utilisateurs novices ; la mise en œuvre de telles connaissances pourrait faciliter la « compréhension », par le système, des intentions exprimées plus ou moins maladroitement par les novices, et permettre à celui-ci de les assister efficacement dans leur réalisation. La compréhension, dans ce contexte, se réduit à la mise en correspondance des effets et résultats escomptés par le novice avec les informations procédurales susceptibles de faciliter leur obtention.

Si l'on parvient à doter les systèmes d'aide en ligne de stratégies contextuelles, il restera à déterminer l'efficacité réelle de cette approche.

Nous avons réalisé une étude expérimentale qui contribue à élucider ce point (voir Capobianco, 2002 et 2003). Elle est présentée dans la section suivante.

### *III.3.3. Evaluation expérimentale de l'efficacité d'une aide en ligne contextuelle*

Notre contribution à l'évaluation ergonomique des aides en ligne contextuelle a pris la forme d'une étude expérimentale impliquant des utilisateurs potentiels. Cette étude porte sur la comparaison de l'efficacité de deux systèmes d'aide à l'utilisation d'un logiciel grand public (Word), l'un contextuel, l'autre non contextuel. Le système contextuel reproduit l'expertise humaine mise en évidence grâce à l'analyse de dialogues expert-novice d'assistance à l'utilisation de Word (voir la section I.2.2).

*Protocole expérimental*

La technique du magicien d'Oz a été utilisée pour simuler partiellement les deux systèmes. Le rôle du compère était limité à l'interprétation des demandes d'aide exprimées oralement [11] par les sujets, et à leur traduction sous la forme de requêtes d'accès à des messages d'aide pré-définis, stockés dans deux bases de données, l'une pour l'aide contextuelle, l'autre pour l'aide non contextuelle. Les messages, pages HTML combinant texte et graphique, étaient envoyés aux sujets via le réseau local. Le poste de travail des sujets comprenait deux écrans côte à côte, l'un dédié aux affichages de Word, l'autre aux messages d'aide. Dix huit étudiants en biologie (1$^{ère}$ année de DEUG) volontaires, qui avaient suivi une formation à Excel mais ne connaissaient pas Word, ont utilisé chacun des deux systèmes d'aide (ordre balancé) pour réaliser dix huit tâches de mise en page de complexité croissante [12], neuf tâches par condition.

Les différences entre messages contextuels et non contextuels étaient les suivantes pour une procédure/tâche quelconque.
- Assistance non contextuelle : Un seul message par procédure/tâche ne contenant aucune référence au contexte d'interaction.
- Assistance contextuelle : Découpage de la procédure/tâche en actions simples, et association d'un message contextuel à chaque action simple. Un message contextuel contenait une ou plusieurs références au contexte d'interaction. A toute demande de description de l'ensemble de la procédure, le système répondait en envoyant le premier message et attendait que la première action ait été exécutée avec succès pour envoyer le second, et continuait ainsi jusqu'au dernier message.

*Résultats*

La comparaison entre les performances des sujets dans les deux conditions s'appuie sur les mesures quantitatives suivantes : pour chaque sujet, le pourcentage de tâches réussies et le pourcentage de tâches réalisées de façon optimum, c'est-à-dire en utilisant la procédure la plus efficace, qui exclut le recours à une approche par essai-erreur notamment.

Le tableau 4 synthétise, pour l'ensemble des tâches et des sujets [13], les valeurs (en pourcentages) des deux premières mesures en fonction du type d'aide utilisé : contextuel (C), non contextuel (NC) et sans recours à l'aide (NA) ; la catégorie NA correspond aux tâches que certains sujets ont exécutées sans faire appel à l'aide en ligne, en procédant par essai-erreur ou en déduisant de la réalisation de tâches antérieures la procédure à appliquer [14]. Pour l'analyse statistique des comparaisons deux à deux des différents ensembles de mesures, des t-tests ont été utilisés.

---
[11] Aucune contrainte d'expression n'était imposée aux sujets qui disposaient d'un microphone. Leurs énoncés étaient transmis au compère via le réseau local.
[12] Session unique d'une durée comprise entre une demi heure et trois quarts d'heure.
[13] Les résultats portent sur 16 sujets seulement, en raison de l'abandon de deux sujets suite à des échecs répétés.

| Mesures | Aide contextuelle | Aide non contextuelle | Sans aide | Analyse statistique | |
|---|---|---|---|---|---|
| Succès (%) | 92 | 81 | 56 | NA vs NC: | **p=0.00206** |
| | | | | NA vs C: | **p=0.00025** |
| | | | | NC vs C: | **p=0.04185** |
| Réalisation optimale (%) | 65 | 59 | 35 | NA vs NC: | **p=0.002206** |
| | | | | NA vs C: | **p=0.000347** |
| | | | | NC vs C: | *p=0.217367* |

Tableau 4 : Performances de l'ensemble des sujets par type d'aide : contextuelle, non contextuelle et sans aide. (Capobianco, 2002).

>   NA : sans aide, C : aide contextuelle, NC : aide non contextuelle.
>   Les pourcentages ont été calculées sur l'ensemble des tâches réalisées par l'ensemble des sujets avec l'assistance du même système d'aide (C ou NC) ou sans aide (NA).
>   Les résultats statistiquement significatifs (t-tests) sont en gras, et les tendances en italique.

A la lecture de ce tableau, on constate que, en accord avec nos attentes, les performances des sujets se sont avérées nettement inférieures, lorsqu'ils ont ignoré le système d'aide en ligne à leur disposition, par rapport à celles qu'ils ont obtenues lorsqu'ils l'ont utilisé. Cependant, contrairement à nos hypothèses de travail, la différence d'efficacité entre les deux systèmes d'aide, contextuel et non contextuel, est peu marquée. Le fait que la différence entre les taux de réussite C et NC soit significative, alors que celle entre les taux de réalisations optimales représente une simple tendance tient vraisemblablement à l'identité des contenus informationnels des messages des deux systèmes d'aide. La faible différence observée entre les deux taux de réussite (C et NC) est, en revanche, plus difficile à expliquer. On trouvera dans Capobianco et Carbonell, 2003) une tentative d'interprétation de cette observation fondée sur l'analyse des résultats des participants à deux tests utilisés couramment en psychologie différentielle :

− Le test GEFT [15] de dépendance-indépendance à l'égard du champ, qui permet d'évaluer la capacité de percevoir un objet particulier dans une scène visuelle, de l'isoler de son contexte. Ce test a été utilisé en interaction homme-machine pour mesurer l'aptitude à explorer les capacités d'un nouveau logiciel, à découvrir les fonctions que celui-ci met à la disposition des utilisateurs (Coventry, 1989 ; Dufresne et Turcotte, 1997).
− Le test BLS4, utile pour évaluer l'ensemble des aptitudes qui interviennent dans la plupart des activités cognitives et que l'on regroupe classiquement sous le terme « intelligence » (Carroll, 1993). La compréhension des messages d'aide et leur interprétation en termes des actions à exécuter et de leur enchaînement mettent en jeu certaines de ces aptitudes.

---

[14] Sur l'ensemble des tâches effectuées (par l'ensemble des sujets), 34% l'ont été sans aide, 32% en faisant appel à l'aide contextuelle, et 34% en faisant appel à l'aide non contextuelle.
[15] Group Embedded Figures Test.

*III.3.4. Conclusions*

Pour l'instant, on ne peut conclure à l'efficacité supérieure des aides en ligne contextuelles par rapport aux aides classiques, pour résoudre les problèmes spécifiques que pose la conception des systèmes d'aide à l'utilisation des logiciels graphiques actuels destinés au grand public. Les prototypes de recherche qui tentent de résoudre ces problèmes ne parviennent pas à satisfaire les critères classiques de qualité ergonomique. Quant aux systèmes commercialisés, ils frustrent les attentes des utilisateurs novices et ne sont que rarement consultés.

D'une part, l'identification dynamique des intentions et buts courants des utilisateurs novices, des informations dont ils ont besoin et du moment opportun pour les leur fournir apparaît, au moins aujourd'hui, comme un obstacle à la réalisation de systèmes d'aide en ligne à l'intention du grand public qui s'adaptent à chaque utilisateur de manière suffisamment flexible et qui soient d'utilisation suffisamment simple pour être effectivement consultés.

D'autre part, il n'est pas certain que les stratégies des experts humains soient un modèle pertinent pour résoudre ces problèmes. L'évaluation ergonomique d'un système simulant leur mise en œuvre et la comparaison de son efficacité avec celle d'un système non contextuel ont fourni des résultats mitigés. Certes, les deux systèmes d'aide ont été largement utilisés, puisque la consultation de l'un ou l'autre système est intervenue dans la réalisation de deux tiers des tâches. En outre, les performances des sujets, lorsqu'ils ne font pas appel à l'aide sont très inférieures [16] à celles qu'ils obtiennent lorsqu'ils consultent l'aide. Cependant, lorsque l'aide est utilisée, les performances des sujets varient peu d'un système d'aide à l'autre.

Des recherches empiriques et expérimentales supplémentaires sont nécessaires pour déterminer l'influence exacte des capacités cognitives individuelles sur l'efficacité des aides en ligne contextuelles et, plus généralement, pour confirmer la pertinence des stratégies d'aide des experts humains. En effet, les utilisateurs interagissent différemment avec un système informatique qu'avec un opérateur humain, et leurs attentes ne sont pas les mêmes. C'est du moins ce que suggèrent les conclusions de l'étude présentée dans (Amalberti et al., 1993). En outre, il n'est pas certain que les techniques d'intelligence artificielle aient atteint une maturité suffisante pour qu'il soit possible, à court terme, d'implémenter les stratégies contextuelles des experts humains de manière satisfaisante.

D'autres directions de recherche méritent également d'être explorées en vue d'éliminer les obstacles spécifiques auxquels sont confrontés les concepteurs d'aides en ligne (voir l'introduction de la section III). Le recours à d'autres modalités que celles utilisées actuellement pour interagir avec les aides en ligne pourrait contribuer à résoudre certains des problèmes spécifiques auxquels sont confrontés les concepteurs de systèmes d'aide.

Nous étudions, dans la section suivante, les possibilités offertes par de nouvelles formes de multimodalité pour améliorer, à moyen terme, l'efficacité des aides en ligne de manière significative.

## III.4. Apports de la multimodalité

La parole en entrée et la voix de synthèse en sortie peuvent être utilisées en parallèle des modalités d'interaction classiques, puisque ces dernières mobilisent essentiellement la perception visuelle et la motricité manuelle, tandis que la parole sollicite exclusivement la perception auditive et que les gestes articulatoires impliqués dans sa production laissent les mains de l'utilisateur totalement libres. L'intégration de la parole, en entrée et en sortie, aux interfaces graphiques actuelles peut donc être mise à profit pour résoudre certains des problèmes spécifiques soulevés par la conception des systèmes d'aide en ligne.

---

[16] Concernant le taux de tâches réussies, on constate une chute de -36% et -25% par rapport aux tâches exécutées avec recours à l'aide contextuelle et à l'aide non contextuelle, respectivement. Concernant les taux de réalisation optimale des tâches, la différence est respectivement de -30% et -29%. Ces données quantitatives sont extraites de (Capobianco, 2002).

Cette section évoque et discute les contributions potentielles de la parole à l'amélioration de l'efficacité des systèmes d'aide en ligne.

### *III.4.1. Contributions potentielles de la parole à l'efficacité des aides en ligne*

Offrir à l'utilisateur les moyens de consulter oralement l'aide, et lui fournir les informations procédurales qu'il sollicite sous forme orale, permettent de résoudre un des trois problèmes spécifiques de conception évoqués dans la section II (voir la sous-section II.2), à savoir :

– l'obligation, pour l'utilisateur, d'interrompre ses interactions avec le logiciel pour consulter l'aide ; il peut interagir oralement avec l'aide tout en poursuivant son interaction gestuelle avec le logiciel ;
– conjointement, le changement de contexte que lui imposent les systèmes d'aide classiques ; la présentation orale des instructions dispense de l'affichage de messages d'aide textuels et/ou graphiques dans une fenêtre dédiée qui recouvre en partie ou entièrement l'affichage courant du logiciel et distrait l'utilisateur de son interaction avec celui-ci.

Cependant, les instructions orales, pour être mémorisées, doivent être courtes. Ce résultat est connu depuis longtemps. Nous avons eu l'occasion de le vérifier dans le cadre d'une étude expérimentale portant sur l'assistance à la recherche de cibles visuelles. Les sujets devaient chercher dans un affichage complexe un élément qui leur avait été décrit auparavant oralement. Pour les assister dans la recherche de la cible, des informations sur sa localisation dans l'affichage étaient incluses dans la description. Nous avons pu constater que les messages longs ne leur ont été d'aucune utilité (Carbonell et Kieffer, 2005). En outre, si les instructions orales sont longues, leur répétition intégrale peut être fastidieuse pour l'utilisateur, surtout si celui-ci souhaite seulement ré-entendre une information unique, ponctuelle. Enfin, alors que l'on peut remplacer avantageusement [17] les textes inclus dans les messages d'aide actuels par des énoncés de synthèse ou pré-enregistrés, la substitution de descriptions orales aux graphiques (copies d'écrans notamment) que comprennent souvent ces messages est impossible, en raison des caractéristiques expressives radicalement différentes de ces deux modalités.

Ces limites intrinsèques de la modalité orale peuvent toutefois être facilement surmontées. Les propositions présentées dans la suite de cette section permettent de s'en affranchir et d'envisager la conception de messages d'aide multimodaux qui mettent à profit les ressources expressives spécifiques de la parole et du graphique sans pour autant imposer à l'utilisateur un changement de contexte lors de la consultation de l'aide. La section se termine par un aperçu des bénéfices que l'on peut escompter de la consultation orale ou multimodale des aides en ligne, moyennant la résolution des problèmes de reconnaissance posés par la mise en œuvre de la parole en entrée.

#### *Conception de messages d'aide multimodaux efficaces*

Pour obtenir des messages oraux courts, donc faciles à comprendre et à mémoriser, il suffit de décomposer la présentation des instructions nécessaires à l'exécution d'une procédure complexe du logiciel en messages courts correspondant chacun à une étape du déroulement de la procédure. Cette stratégie est facile à mettre en œuvre. Elle semble en outre efficace et bien acceptée par les utilisateurs novices du grand public. C'est du moins ce que suggèrent les performances et les jugements subjectifs des 18 sujets qui ont participé à l'étude expérimentale présentée dans la section III.3.3 (paragraphe 'Protocole') où nous l'avons utilisée pour élaborer les messages textuels du système d'aide contextuel. A noter qu'elle est indépendante de la stratégie d'aide adoptée, contextuelle ou non. Simplement, dans un contexte d'aide non contextuelle, l'activation successive des instructions qui composent la description d'une procédure complexe sera nécessairement à la charge de l'utilisateur ; elle ne pourra être assurée par le système car celui-ci manquera des informations contextuelles indispensables pour prendre des décisions pertinentes concernant le choix du moment opportun pour émettre chacun des différents messages associés à la réalisation d'une même procédure.

---

[17] En particulier, si la lecture de textes affichés de plusieurs lignes constitue une activité inhabituelle pour l'utilisateur, elle requiert de sa part un effort spécifique qui augmente sa charge de travail mental. Cet effort supplémentaire est susceptible de nuire à la compréhension et à la mémorisation des messages d'aide textuels.

Quant à la seconde limite mentionnée, deux solutions simples sont envisageables pour la surmonter. Toutes deux conservent la modalité graphique, irremplaçable, mais éliminent son inconvénient majeur, le recouvrement partiel ou total des affichages du logiciel par la fenêtre d'aide, qui est perçu par l'utilisateur comme une intrusion.

La première solution consiste à utiliser deux écrans, un pour l'interaction avec le logiciel, l'autre pour l'affichage des messages d'aide. Les multi-écrans se banalisent aujourd'hui [18], ce qui rend cette solution abordable. C'est celle que nous avons utilisée dans le cadre de l'étude expérimentale des stratégies d'aide contextuelle et non contextuelle (voir la section III.3.3). Elle a donné entière satisfaction aux sujets. Une variante de cette solution consiste à projeter côte à côte, sur un écran de grandes dimensions, les deux affichages. Mais cette variante présente plusieurs inconvénients dont les principaux sont, outre un certain dépaysement créé par la distance et les dimensions inhabituelles de l'écran, la grande luminosité des affichage source éventuelle de fatigue visuelle, et la nécessaire pénombre de l'environnement, susceptible de perturber les interactions que l'utilisateur est amené à effectuer avec son environnement, parallèlement à son interaction avec le logiciel (e.g., consultation de documents papier).

La seconde solution, la plus efficace en raison de sa simplicité et la plus naturelle, consiste à « montrer » à l'utilisateur comment faire pour réaliser une de ses intentions, par exemple une intention dont la réalisation implique l'exécution d'une procédure complexe. C'est-à-dire que le système désigne sur l'affichage courant du logiciel les objets graphiques évoqués dans une instruction, en même temps qu'il les mentionne oralement lors de l'énoncé de cette instruction.

La langue naturelle permet en effet au locuteur d'évoquer dans ses énoncés les objets du monde physique qui l'entourent, grâce aux expressions déictiques [19]. Ces expressions font référence à l'environnement physique du locuteur ; celui-ci accompagne éventuellement leur énoncé d'un geste de désignation, notamment lorsque la référence est ambiguë. Au cas particulier, l'environnement partagé par le locuteur, à savoir le système, et par l'interlocuteur, à savoir l'utilisateur, est l'affichage courant du logiciel. Quant au geste de désignation, il peut être simulé par n'importe quelle mise en relief visuelle de l'objet graphique présent à l'écran auquel le système fait référence dans l'instruction en cours d'énonciation. Une autre solution, peu différente, est la « démonstration » des actions à effectuer sur l'interface pour exécuter une procédure. C'est celle qui est mise en œuvre par l'*Office Assistant* de Microsoft qui propose à l'utilisateur d'exécuter à sa place les actions nécessaires à la réalisation d'une procédure ayant fait l'objet d'une demande d'informations de la part de celui-ci. En associant à la démonstration un commentaire oral partiellement redondant, on facilite la mémorisation des instructions. Toutefois, en l'absence de commentaire oral, l'efficacité de cette approche est discutable. En effet, le novice risque d'éprouver des difficultés à assimiler et mémoriser un enchaînement d'actions qu'il n'aura pas effectuées lui-même, qu'il aura seulement vu se dérouler devant lui, rapidement et sans aucun commentaire verbal. C'est du moins le jugement que sont susceptibles de porter sur cette innovation les partisans de l'apprentissage par l'action.

---

[18] A côté des solutions intégrées commercialisées, des logiciels libres permettent d'utiliser simultanément deux moniteurs classiques sous Unix. Voir, par exemple, pour la première catégorie de solutions le site de 9X Media (http:www.9xmedia.com) et, pour la seconde, la page d'adresse :
    http:www.sismo.ethz.ch/linux/xinerama.html

[19] « ici/là », « je/tu/nous/vous» sont les déictiques les plus usités. Ils font référence, respectivement, au lieu où se déroulent les échanges verbaux et aux partenaires de ces échanges (locuteur et interlocuteur(s)).

*Apports de la parole en tant que modalité de consultation de l'aide*

La possibilité de consulter l'aide oralement offre *a priori* plusieurs avantages, au moins pour les utilisateurs novices et occasionnels issus du grand public.

D'abord, il convient de rappeler, pour mémoire, l'intérêt majeur que présente le recours à la parole pour interagir avec l'aide. Elle est la modalité d'entrée qui réduit au maximum les interférences entre l'activité principale de l'utilisateur, son interaction avec le logiciel, et la consultation de l'aide.
D'autre part, l'utilisation de la parole dans ce contexte place l'utilisateur dans une situation familière pour lui, la demande de renseignements à distance, par téléphone. Cette analogie, en permettant le transfert de compétences antérieures à l'activité nouvelle que représente la consultation de l'aide pour l'utilisateur novice, est de nature à rendre inutile tout apprentissage de l'utilisation d'une aide en ligne quelconque. Le novice ne dispose aujourd'hui d'aucune opportunité de ce type, car l'accès aux systèmes d'aide actuels requiert une familiarité avec des techniques de consultation que certains utilisateurs peuvent ignorer ou mal maîtriser, comme l'extraction d'informations par sélection de mots-clés, la navigation à l'aide de liens hypertextes ou le parcours d'une « table des matières électronique » organisée hiérarchiquement. Même si ces techniques lui sont familières, leur mise en œuvre a un coût cognitif non négligeable, dans la mesure où il doit distraire temporairement son attention de l'interaction avec le logiciel.
Plus généralement, en donnant à l'utilisateur les moyens de consulter l'aide oralement, on contribue de façon significative à la résolution des trois problèmes spécifiques auxquels sont confrontés les concepteurs d'aides en ligne, à condition toutefois que celui-ci puisse s'exprimer librement, sans contrainte linguistique majeure, et que la reconnaissance de ses requêtes soit suffisamment fiable. La langue naturelle orale est en effet considérée comme le mode de communication le plus « naturel » et le plus utilisé. Sa mise en œuvre en tant que mode d'accès à l'aide permet de prendre en compte, implicitement, la grande diversité cognitive de la vaste population d'utilisateurs formant le grand public. Si les réponses du système aux requêtes du novice utilisent la terminologie courante et réduisent, en les explicitant ou en les expliquant, les divergences de points de vue entre le concepteur et le grand public sur le domaine d'application et les tâches que le logiciel permet de réaliser, le premier défi de conception (voir la section II.1) pourra être considéré comme résolu en grande partie. Le parallélisme possible entre l'interaction avec le logiciel et la consultation de l'aide qu'autorise l'usage de la parole pour exprimer les demandes d'informations d'aide devrait permettre de progresser dans la résolution du second défi de conception (voir la section II.2), à condition de laisser l'initiative des échanges à l'utilisateur et de réduire le dialogue entre celui-ci et le système d'aide à une suite de couples question-réponse. Toute intervention à l'initiative de l'aide risque en effet d'être perçue par l'utilisateur comme une interruption intempestive, et un dialogue complexe comme une distraction susceptible d'interférer avec son activité principale. Quant au troisième et dernier défi de conception, qui résulte du « paradoxe de motivation » (voir la section II.3), on peut considérer que la solution proposée contribue à le résoudre partiellement. En effet, dans la mesure où on permet un accès aux informations d'aide plus aisé, qui présente un coût cognitif faible et interrompt au minimum l'utilisateur dans son activité principale, on peut espérer que celui-ci consultera l'aide plus souvent à condition, bien sûr, que les informations en réponse à ses requêtes satisfassent ses besoins et lui soient intelligibles.

Cette voie de recherche a été considérée et abordée très tôt, mais les efforts engagés se sont avérés infructueux. Des tentatives dont l'objectif à long terme était la mise en œuvre de dialogues d'aide en langue naturelle ont vu le jour dès le début des années 90 (Pilkington 1992 ; Roestler et McLellan, 1995). Mais elles n'ont pas été poursuivies, peut-être parce que les dialogues envisagés étaient complexes, et les échanges textuels. Les résultats de l'étude expérimentale présentée dans la section III.3.3, en revanche, incitent à poursuivre l'exploration de la direction de recherche proposée ci-dessus, qui diffère des approches antérieures, dans la mesure où elle préconise simplement l'utilisation de la parole comme modalité de consultation de l'aide dans un contexte d'échanges de type question-réponse, plutôt que la mise en œuvre de véritables dialogues d'aide dont la fiabilité et la robustesse risquent de s'avérer insuffisantes. A titre de rappel, les dix huit sujets qui ont participé à l'expérience pouvaient formuler leurs demandes d'aide oralement, en l'absence de toute contrainte d'expression. Or, ce à quoi ils ont été le plus sensibles dans leurs jugements subjectifs, ce ne sont ni les avantages ni les inconvénients des deux systèmes d'aide qu'ils étaient censés évaluer, mais la possibilité de consulter ces systèmes oralement. En effet, un tiers d'entre eux ont mentionné spontanément que cette possibilité était ce qu'ils avaient le plus apprécié dans l'interaction avec l'aide. Ils ont exprimé cette appréciation, parfois avec enthousiasme, soit, à la suite de l'expérimentation, dans la rubrique 'Autres commentaires' du questionnaire d'évaluation des deux systèmes d'aide, soit, ultérieurement, pendant l'entretien semi-directif final (*debriefing*). Ce jugement est à mettre en relation avec le fait que la réalisation de deux tiers des tâches (réussies ou non) effectuées par l'ensemble des sujets a fait l'objet de recours à l'aide.

Cependant, avant de poursuivre l'exploration de cette direction de recherche et d'entreprendre la réalisation de systèmes d'aide en ligne consultables oralement, il est nécessaire de s'assurer que les performances des systèmes actuels de reconnaissance de la parole et d'interprétation de la langue naturelle sont suffisantes pour traiter les requêtes orales des utilisateurs de manière fiable sans imposer de contraintes artificielles sur leur expression spontanée. Ces deux conditions, fiabilité de la reconnaissance-interprétation et absence de contraintes d'expression artificielles, doivent être remplies pour que la parole en tant que modalité d'accès aux informations d'aide obtienne l'adhésion des utilisateurs, notamment celle du grand public.

Des travaux antérieurs sur la conception et l'évaluation ergonomique de langages d'interaction oraux « acceptables » contribuent à rendre réaliste la voie de recherche proposée. Nous avons montré, en effet, qu'il était possible de donner aux utilisateurs des moyens d'expression orale qui permettent une interaction efficace grâce à une reconnaissance robuste de leurs énoncés par les systèmes actuels, tout en étant bien acceptés par eux dans la mesure où ils n'imposent à leur expression spontanée que des contraintes légères, faciles à acquérir au cours de l'interaction, et ont un pouvoir d'expression suffisant, contrairement aux langages artificiels classiques dont le pouvoir d'expression est limité, voire insuffisant, et dont la maîtrise nécessite un apprentissage spécifique, préliminaire à leur utilisation. La méthode que nous proposons pour définir de tels langages est décrite dans (Carbonell, 2001 et 2003), et les résultats de l'évaluation ergonomique expérimentale d'un langage oral d'interaction défini à l'aide de cette méthode sont présentés dans (Robbe et al. 2000) ; cette évaluation a impliqué 12 participants, utilisateurs potentiels de tels langages.

## IV CONCLUSIONS

La conception d'aides en ligne à l'intention du grand public soulève des problèmes spécifiques difficiles à résoudre. Pour qu'un système d'aide en ligne soit effectivement consulté par les utilisateurs non informaticiens, novices ou occasionnels, les concepteurs doivent surmonter trois obstacles majeurs. Le premier obstacle tient à l'hétérogénéité de cette population, à la diversité des profils d'utilisateurs dont il faut tenir compte. La diversité porte notamment sur les capacités cognitives, les savoirs, savoir-faire et expériences antérieures, les intérêts, les préférences et les attitudes des utilisateurs ciblés. Le second est lié à l'interruption de l'interaction avec le logiciel que nécessite toute consultation de l'aide. Le dernier concerne l'attitude spécifique des utilisateurs non informaticiens et non professionnels, en particulier des novices, vis-à-vis des logiciels d'aide en ligne. Cette attitude résulte de leurs motivations dont les particularités constituent, selon les chercheurs qui les ont mises en évidence, le « paradoxe de motivation ». Les techniques développées récemment dans le domaine de la conception des interfaces utilisateur ne semblent pas en mesure, au moins à court terme, de résoudre ces problèmes. Ces techniques incluent, notamment, l'adaptation statique ou dynamique de l'interaction aux caractéristiques spécifiques de l'utilisateur courant, l'application des principes de conception universelle et, dans le domaine de l'aide en ligne, celle de stratégies d'aide contextuelles. La mise en œuvre de ces différents concepts et techniques et/ou la qualité ergonomique des prototypes résultants s'avèrent, au moins pour l'instant, décevantes.

Une voie de recherche prometteuse à court terme est possible dans l'attente des progrès scientifiques importants, nécessaires pour assurer la viabilité des approches évoquées précédemment : la multimodalité. Cette dernière offre des perspectives de solution aux problèmes spécifiques de conception des aides en ligne, certes partielles, mais dont la mise en œuvre est envisageable dès à présent. En effet, en donnant à l'utilisateur les moyens de consulter oralement le système d'aide en ligne, on réduit sensiblement les effets négatifs de l'interruption nécessaire de son activité principale, l'interaction avec le logiciel. En outre, si le novice a la possibilité d'exprimer ses demandes d'informations en langue naturelle, on peut supposer que l'aide en ligne sera vraisemblablement consultée beaucoup plus souvent qu'elle ne l'est actuellement, car le manque de transparence de l'interaction avec les systèmes existants est un frein certain à leur utilisation. C'est ce que suggèrent les comportements de sujets (18) qui ont expérimenté deux systèmes d'aide simulés qu'ils pouvaient interroger oralement en l'absence de toute contrainte d'expression.

Cependant, la fiabilité et la robustesse des systèmes actuels de reconnaissance de la parole ne sont pas suffisantes pour qu'il soit possible, à court terme, d'offrir à l'utilisateur une interaction orale totalement transparente. Toutefois, on est en mesure, dès maintenant, d'envisager la définition et la mise en œuvre, pour la consultation des aides en ligne, de langages multimodaux « acceptables », qui soient susceptibles de recueillir l'adhésion des utilisateurs novices et occasionnels issus du grand public car ils imposent un contrôle de l'expression spontanée limité à un ensemble réduit de contraintes faciles à maîtriser sans apprentissage spécifique.

Quant à la mise en œuvre de la parole comme modalité d'expression des informations d'aide, elle présente des difficultés d'ordre ergonomique plutôt que technique, solubles à court ou moyen terme. Les deux problèmes principaux à résoudre peuvent se résumer ainsi. Quels moyens de désignation visuelle (des objets graphiques affichés à l'écran) donner au système d'aide pour lui permettre une utilisation des déictiques efficace, qui n'interfère pas avec l'activité principale de l'utilisateur, à savoir son interaction avec le logiciel ? Comment limiter, sans engager de véritable dialogue avec le système d'aide, l'inconvénient majeur des messages oraux par rapport aux messages textuels qui est d'obliger l'utilisateur à écouter ou ré-écouter l'intégralité d'un, voire plusieurs messages, lorsqu'il recherche une information ponctuelle, par exemple le rôle d'un paramètre d'une boîte de dialogue ? Nous avons pu constater, dans le cadre d'une expérimentation (en voie d'achèvement) centrée sur l'évaluation d'un système d'aide oral à l'utilisation de Flash par une vingtaine de participants, que cet inconvénient de la modalité orale par rapport au texte est le seul qui ait été signalé dans les formulaires d'évaluation comme lors des entretiens de *debriefing*.